\documentclass[prl,twocolumn,letterpaper,10pt,twoside,tightenlines,nofootinbib,showpacs]{revtex4}
\usepackage{amsmath,amsfonts,amssymb,latexsym}
\usepackage{graphicx,color}
\usepackage[sort&compress]{natbib}

\begin{document}

\newcommand{\eg}{{\it e.g.}}
\newcommand{\cf}{{\it cf.}}
\newcommand{\etal}{{\it et. al.}}
\newcommand{\ie}{{\it i.e.}}
\newcommand{\be}{\begin{equation}}
\newcommand{\ee}{\end{equation}}
\newcommand{\bea}{\begin{eqnarray}}
\newcommand{\eea}{\end{eqnarray}}
\newcommand{\bef}{\begin{figure}}
\newcommand{\eef}{\end{figure}}
\newcommand{\bce}{\begin{center}}
\newcommand{\ece}{\end{center}}
\newcommand{\red}[1]{\textcolor{red}{#1}}

\newcommand{\dd}{\text{d}}
\newcommand{\ii}{\text{i}}
\newcommand{\lsim}{\lesssim}
\newcommand{\gsim}{\gtrsim}
\newcommand{\RAA}{R_{\rm AA}}

\title{Hadronization and Charm-Hadron Ratios in Heavy-Ion Collisions}

\author{Min~He$^{1}$ and Ralf~Rapp$^2$}
\affiliation{$^1$Department of Applied Physics, Nanjing University of Science and Technology, Nanjing~210094, China}
\affiliation{$^2$Cyclotron Institute and Department of Physics and
Astronomy, Texas A\&M University, College Station, Texas 77843-3366, U.S.A.}

\date{\today}

\begin{abstract}
Understanding the hadronization of the quark-gluon plasma (QGP) remains a challenging problem in the study of strong-interaction
matter as produced in ultrarelativistic heavy-ion collisions
(URHICs). The large mass of heavy quarks renders them excellent tracers of the color neutralization process of the QGP when they
convert into various heavy-flavor (HF) hadrons. We develop a
4-momentum conserving recombination model for HF mesons and baryons that recovers the thermal and chemical equilibrium limits and accounts for space-momentum correlations (SMCs) of
heavy quarks with partons of the hydrodynamically expanding QGP,
thereby resolving a long-standing problem in quark
coalescence models. The SMCs enhance the recombination of
fast-moving heavy quarks with high-flow thermal quarks in the
outer regions of the fireball. We also improve the
hadro-chemistry with ``missing" charm-baryon states, previously
found to describe the large $\Lambda_c/D^0$  ratio observed in
proton-proton collisions. Both SMCs and hadro-chemistry, as part
of our HF hydro-Langevin-recombination model for the strongly
coupled QGP, importantly figure in the description of recent
data for the $\Lambda_c/D^0$ ratio and $D$-meson elliptic flow
in URHICs.

\end{abstract}


\maketitle

{\it Introduction.---}
Ultra-relativistic heavy-ion collisions (URHICs) at RHIC and the
LHC have created a novel state of strong-interaction matter composed of deconfined quarks and gluons, the
Quark-Gluon Plasma (QGP)~\cite{Akiba:2015jwa,Shuryak:2014zxa}. The QGP behaves like a near-perfect fluid with small specific
shear viscosity, as revealed by the collective flow patterns in final-state hadron spectra being consistent with relativistic
hydrodynamic simulations~\cite{Heinz:2013th,Gale:2013da,Niemi:2015qia}. A closely related
discovery is the surprisingly large collective flow observed for heavy-flavor (HF) particles and requiring a small diffusion coefficient, ${\cal D}_s$~\cite{Rapp:2018qla,Dong:2019byy},
corroborating the strongly-coupled nature of the QGP.
Another interesting finding is an enhancement of baryon-to-meson
ratios ($p/\pi$ and $\Lambda/K$), relative to $pp$ collisions, at intermediate transverse momenta, $p_T$$\simeq$3-4\,GeV,
together with the so-called constituent-quark number scaling (CQNS) of the ellpitic flow, $v_2$, of baryons and mesons.
These observations have been attributed to quark coalescence as
a hadronization mechanism of kinetic (non-thermalized) partons with thermal partons in the
QGP~\cite{Greco:2003xt,Fries:2003vb,Hwa:2003bn,Molnar:2003ff}.
In this paper we will argue that the diffusion and hadronization
of HF particles provide a unique opportunity to put these
phenomena on a common ground.

The diffusion of low-momentum HF particles has long
been recognized as an excellent gauge of their interaction
strength with the medium, most notably through their $v_2$
acquired in non-central URHICs via a drag from the collectively
expanding fireball, \cf~Ref.~\cite{Dong:2019byy} for a recent
review. The large heavy-quark (HQ) mass, $m_Q\gg T_{H}$ (with $T_H$$\simeq$160\,MeV the typical hadronization
temperature~\cite{Andronic:2017pug}), also opens a direct window on hadronization processes.
Thus, HF spectra simultaneously encompass the strong-coupling of the QGP and its hadronization.
In particular, the chemistry of the produced HF hadrons~\cite{Andronic:2007zu,Kuznetsova:2006bh,He:2012df,Oh:2009zj,Plumari:2017ntm},
has recently drawn a lot of attention through the observed
enhancements in the $D_s/D^0$ and $\Lambda_c/D^0$ ratios at
RHIC~\cite{Zhou:2017ikn,Adam:2019hpq} and the
LHC~\cite{Acharya:2018hre,Acharya:2018ckj}. Reliable
interpretations of these data require hadronization models that satisfy both kinetic and chemical equilibrium in the limit of
thermal quark distributions as an input. This is also a
pre-requisite for an ultimate precision extraction of the HF
transport coefficients, reinforcing the intimate relation between HQ diffusion and hadronization.
In the kinetic sector, this has been achieved in the resonance
recombination model (RRM)~\cite{Ravagli:2007xx}, where a
conversion of equilibrium quark- to $D$-meson spectra in URHICs,
including their $v_2$, has been established
on a hydrodynamic hypersurface~\cite{He:2011qa}. As the
RRM is based on resonance correlations that develop near
$T_H$ in heavy-light $T$-matrix interactions~\cite{Riek:2010fk},
it directly connects to a small HQ diffusion coefficient in
the QGP.

In this work, we develop and implement several concepts in quark recombination that will be critical in a comprehensive
set-up for HF phenomenology in URHICs. First, we derive a 4-momentum conserving three-body recombination formula for the
hadronization into baryons, and verify its quark-to-baryon equilibrium mapping.
Second, we devise an event-by-event implementation
for HQ distributions obtained from Langevin simulations, which maintains HQ number conservation and satisfies the equilibrium limit of the HF hadro-{\em chemistry}. The event-by-event HQ number
conservation is pivotal in a precise treatment of
space-momentum correlations (SMCs) of individually transported
heavy quarks with anti-/quarks of the underlying hydro
background. Both hadro-chemistry and quark SMCs have been
challenging issues for instantaneous coalescence models
(ICMs)~\cite{Molnar:2004rr,Fries:2008hs};
as such our developments are
pertinent well beyond the HF sector. Third, the equilibrium
limit of the HF hadro-chemistry is improved by employing a large
set of ``missing" HF baryon states not listed by the particle
data group (PDG), but predicted by the relativistic-quark model (RQM)~\cite{Ebert:2011kk} and consistent with lattice-QCD (lQCD) computations~\cite{Bazavov:2014yba,Padmanath:2014bxa}.
In Ref.~\cite{He:2019tik} they were shown to account for the large $\Lambda_c/D^0$ ratio measured in $pp$ collisions at the LHC (while the environment in $e^+e^-$ collisions is less
conducive to charm-baryon formation).

{\it Baryons in RRM.--}
We first recall the main features of the 2-body
RRM~\cite{Ravagli:2007xx}.
Starting from the Boltzmann equation, resonant quark-antiquark
scattering into mesons near equilibrium,
$q+\bar{q}\leftrightarrow M$,
can be utilized to equate gain and loss terms and arrive at a meson phase space distribution
(PSD) of the form
\begin{align}
f_M(\vec x,\vec p)=\frac{\gamma_{M}(p)}{\Gamma_M}
\int\frac{d^3\vec p_1 d^3\vec p_2}{(2\pi)^3}f_q(\vec x,\vec p_1)f_{\bar q}(\vec x,\vec p_2)
\nonumber\\
\times \sigma_M(s)v_{\rm rel}(\vec p_1,\vec p_2)\delta^3(\vec p -\vec p_1-\vec p_2) \ ,
\label{rrm1}
\end{align}
where $f_{\bar q,q}$ are the anti-/quark PSDs, $v_{\rm rel}$ their relative velocity,
$\gamma_{M}(p)=E_M(p)/m_M$, and $\Gamma_M$ the meson width. The latter, together
with the meson mass $m_M$ and degeneracy factors, also appear
in the resonant $q+\bar{q}\rightarrow M$ cross section, usually
taken of Breit-Wigner type.

The generalization to the 3-body case is conducted
in two steps. First, quark-1 and quark-2 recombine into a diquark,
$q_1(\vec p_1)+q_2(\vec p_2)\rightarrow dq (\vec p_{\rm 12})$, whose
PSD is obtained in analogy to meson formation, by replacing
$M$$\to$$dq$, $q$$\to$$q_1$ and $\bar q$$\to$$q_2$ in
Eq.~(\ref{rrm1}). Diquark configurations are an inevitable component of a thermal QGP approaching hadronization.
Second, the diquark recombines with quark-3 into a baryon
reusing Eq.~(\ref{rrm1}),
\begin{align}
\label{RRM-3}
&f_B(\vec x,\vec p\,)=\frac{\gamma_{B}}{\Gamma_B}
\int\frac{d^3\vec p_1d^3\vec p_2d^3\vec p_3}{(2\pi)^6}\frac{\gamma_{dq}}{\Gamma_{dq}}
f_1(\vec x,\vec p_1)f_2(\vec x,\vec p_2)
\nonumber\\
&\times f_3(\vec x,\vec p_3)\sigma_{dq}(s_{12})v_{\rm rel}^{12}
\sigma_B(s)v_{\rm rel}^{dq3}\delta^3(\vec p -\vec p_1-\vec p_2-\vec p_3) \ ,
\end{align}
where $s_{12} =(p_1+p_2)^2$,  $s=(p_1+p_2+p_3)^2$, $\sigma_{B}$: resonance cross section
for $dq+q\to B$.
This expression depends on the underlying three-quark PSDs on an equal footing.

\begin{figure}[!t]
\hspace{-0.35cm}
\begin{minipage}{0.249\textwidth}
\includegraphics[width=1.18\textwidth]{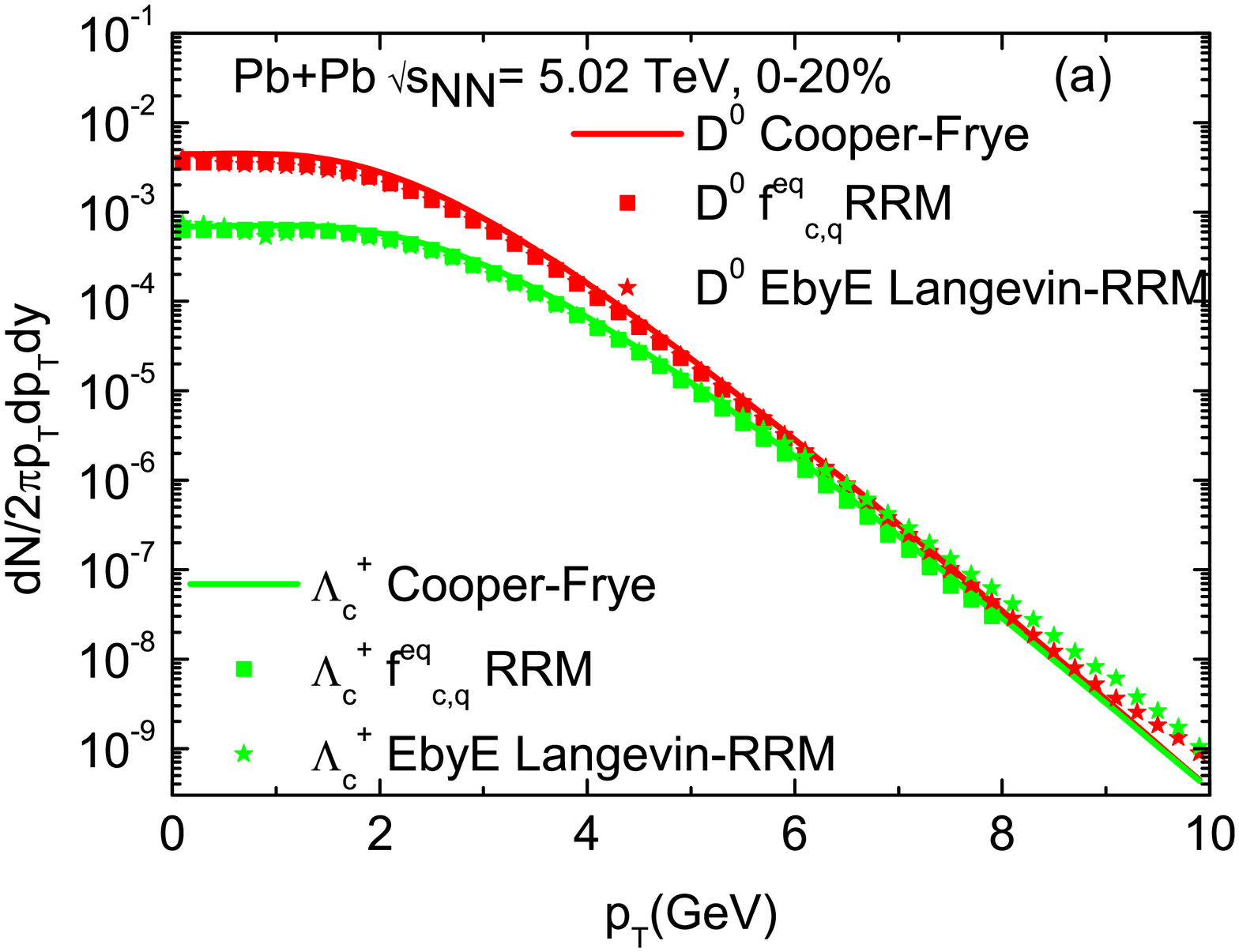}
\end{minipage}
\hspace{-0.22cm}
\begin{minipage}{0.249\textwidth}
\includegraphics[width=1.18\textwidth]{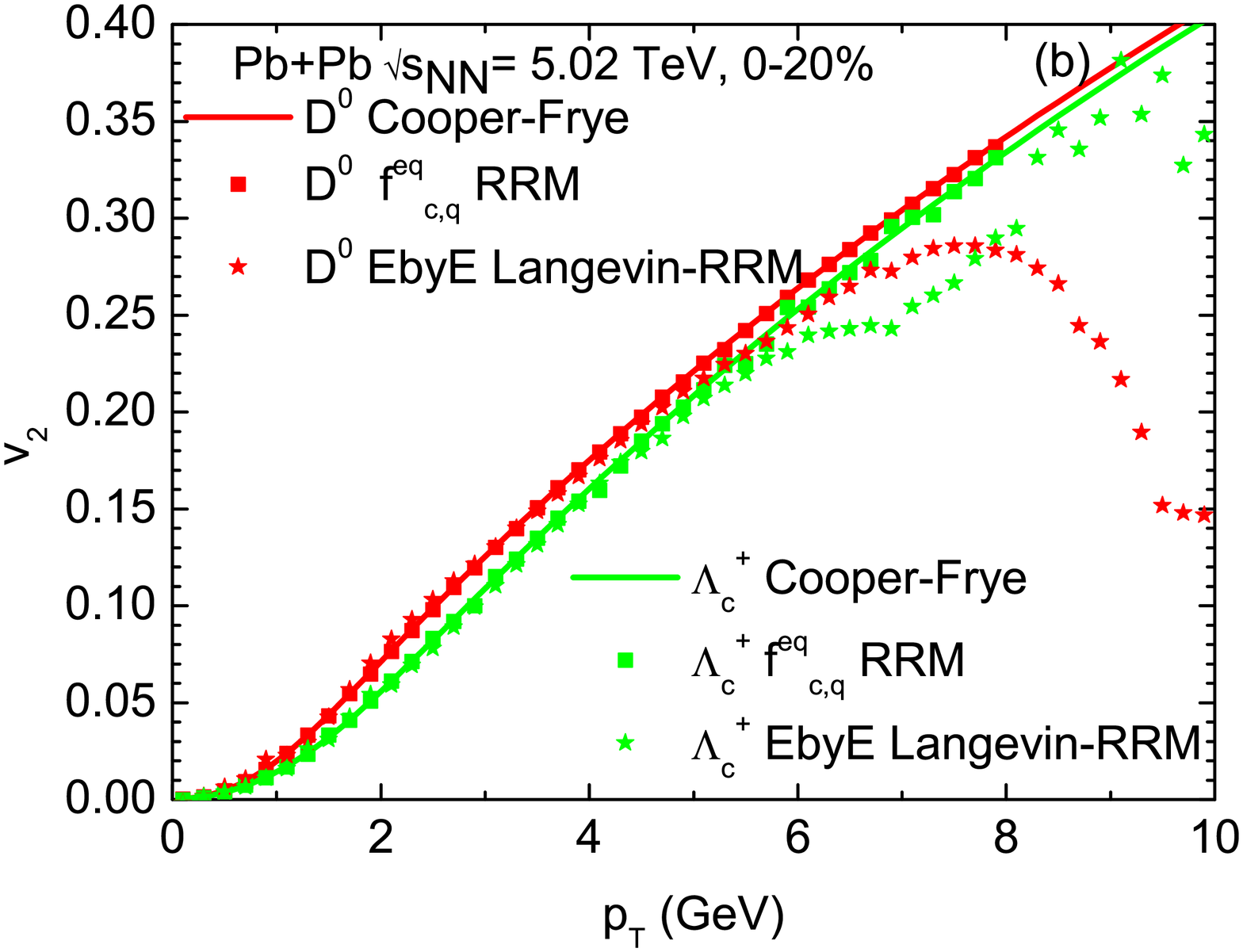}
\end{minipage}
\caption{(Color online) RRM mapping of thermal light- and $c$-quark distributions
(boxes: thermal, stars: from Langevin simulations with large relaxation rate) into (a) $p_T$-spectra
and (b) $v_2$ of $D^0$ and $\Lambda_c^+$, compared to direct $D^0$ and $\Lambda_c^+$ hydro results (lines).}
\label{fig_eq-map}
\end{figure}

To check the equilibrium mapping of quark into hadron spectra,
we calculate the PSDs for recombination of thermal $c$ and light quarks ($q$) into $D^0$ and
$\Lambda_c^+$, using $f_{c,q}^{\rm eq}(\vec x,\vec p)=g_{c,q} e^{-p\cdot u(x)/T_H}$
with a flow velocity $u(x)$ on a hydrodynamic hypersurface at $T_{\rm H}=170$\,MeV for 0-20\%
Pb-Pb ($\sqrt{s_{\rm NN}}$=5.02\,TeV) collisions 
(with quark masses $m_c$=1.5\,GeV, $m_q$=0.3\,GeV and  diquark mass $m_{ud}$=0.7\,GeV).
The invariant hadron spectra,
\begin{align}
\label{dNdp}
\frac{dN_{M,B}}{p_Tdp_Td\phi_pdy}=\int \frac{p\cdot d\sigma}{(2\pi)^3} f_{M,B}(\vec x,\vec p)
\end{align}
($d\sigma_{\mu}$: hypersurface element), displayed in Fig.~\ref{fig_eq-map}, confirm that the
RRM-generated hadron $p_T$-spectra agree with their direct calculation on the same hypersurface in
chemical equilibrium, including their elliptic flow $v_2$, demonstrated here for the first time for baryons.
%

{\it Space-momentum correlations.---}
The original derivation of CQNS for light-hadron $v_2$ within
ICMs assumed spatially homogeneous (global) quark distributions in the fireball,
$v_2^q(\vec x,\vec p)$=$v_2^q(\vec p)$~\cite{Greco:2003xt,Fries:2003vb}.
This is at variance with hydrodynamic flow fields and rendered CQNS to be very fragile upon including SMCs~\cite{Molnar:2004rr}.
The application of RRM for mesons~\cite{Ravagli:2008rt} could
resolve this problem, but no explicit signature of SMCs from recombination processes was
identified (see also Ref.~\cite{Gossiaux:2009mk} using an ICM). Here, we propose that the recent results for the $\Lambda_c/D^0$ ratio, as well as the $p_T$ dependence of
their $v_2$, provide such signatures, and quantitatively
elaborate them within our strongly-coupled
hydro-Langevin approach~\cite{He:2011qa}.

\begin{figure}[!t]
\hspace{-0.35cm}
\begin{minipage}{0.249\textwidth}
\includegraphics[width=1.18\textwidth]{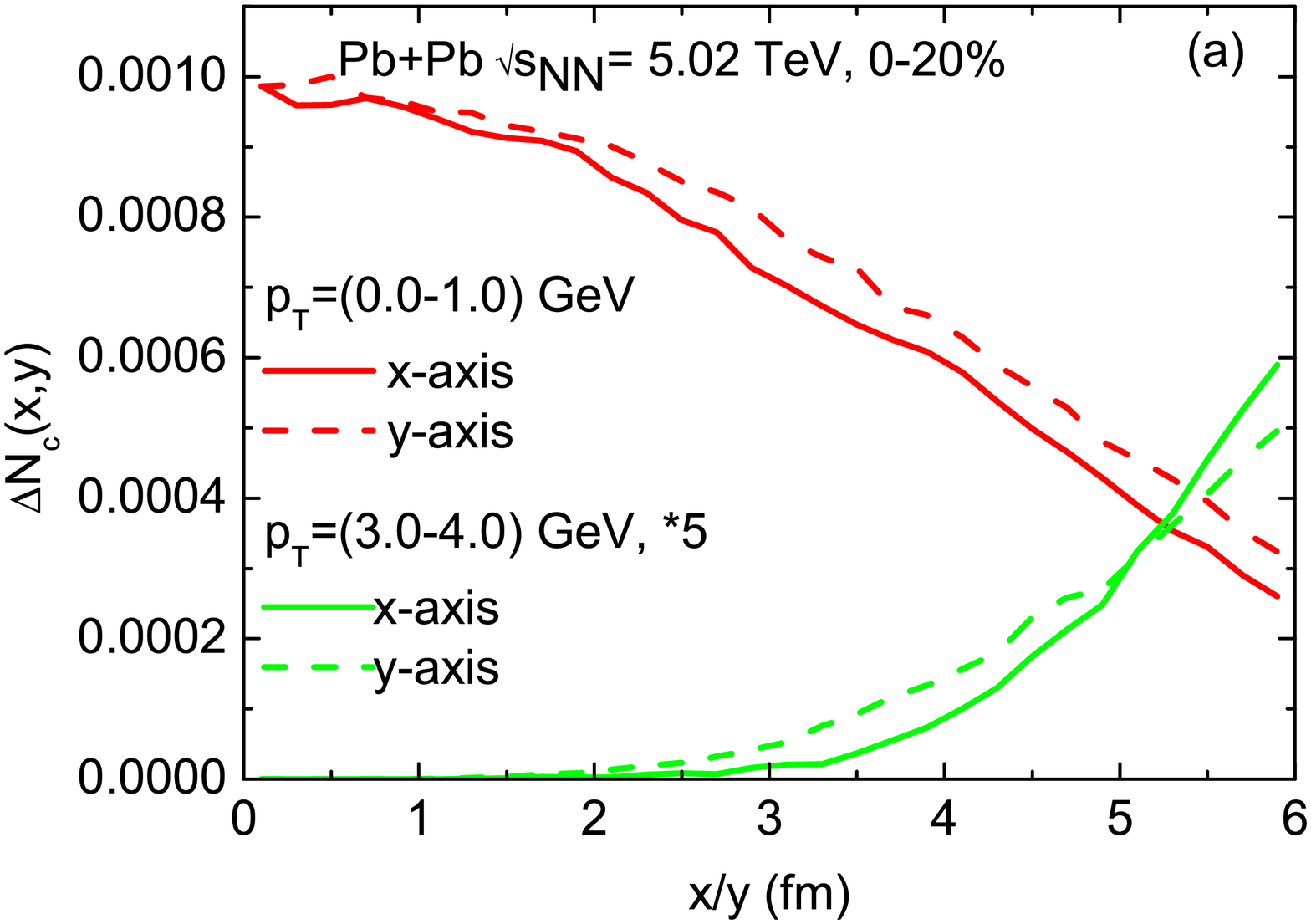}
\end{minipage}
\hspace{-0.22cm}
\begin{minipage}{0.249\textwidth}
\includegraphics[width=1.18\textwidth]{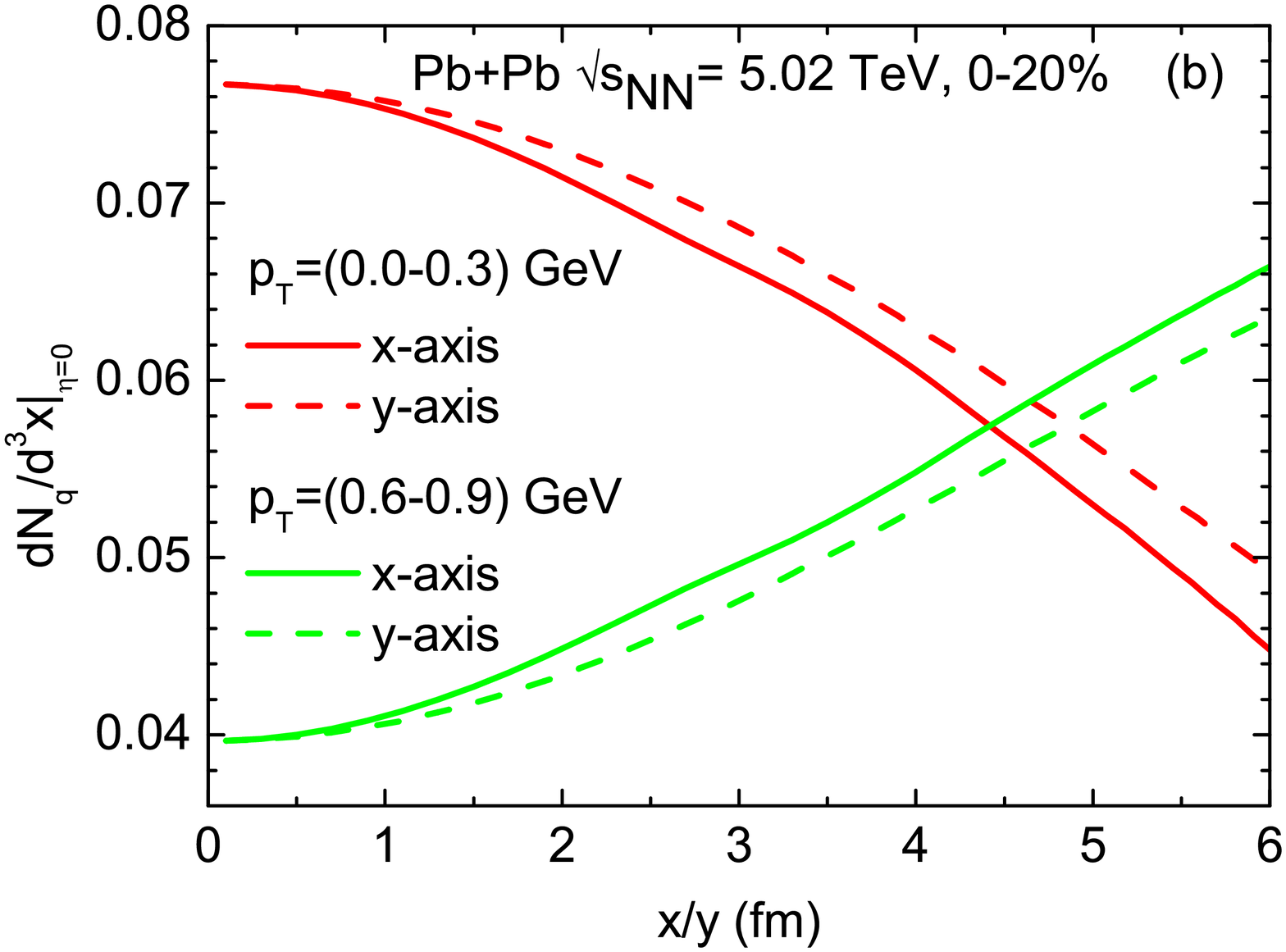}
\end{minipage}
\caption{(Color online) Spatial distributions of (a) $c$-quarks from Langevin simulations and (b) light quarks from hydrodynamics in the transverse fireball plane in various $p_T$ bins.}
\label{fig_smc}
\end{figure}

To begin with, we illustrate the pertinent SMCs in
Fig.~\ref{fig_smc} for $c$-quark distributions in the transverse
plane in different $p_T$ bins at hadronization. Clearly,
low-$p_T$ (0-1\,GeV) and  higher-$p_T$ (3-4\,GeV) $c$ quarks
preferentially populate the inner and outer regions of the
fireball, respectively. The spatial density, $dN/d^3x$, of
Cooper-Frye generated thermal light-quark spectra at midrapidity
from the underlying hydro evolution on the same hypersuface
shows a similar behavior. As recombination occurs between partons close in both $\vec x$ and $\vec p$ space, the
SMCs (not included in studies using
ICMs~\cite{Oh:2009zj,Plumari:2017ntm})
are expected to be relevant in the formation of charm hadrons
especially at intermediate $p_T$ where signals of the baryon enhancement
are prominent.

{\it Event-by-event Langevin-RRM.--}
Since a direct implementation of SMCs with off-equilibrium
$c$-quark PSDs on the full hadronization hypersurface is not
straightforward, we have developed an event-by-event procedure for each
diffusing $c$ quark once it enters a hydro cell at $T_H$. Toward
this end, we first determine the recombination probability,
$P_{M,B}(p_c^*)$, for the $c$-quark as a function of its momentum in the local restframe (starred variables), convert this into hadron PSDs by sampling the thermal light-quark PSDs
at $T_H$, and then evaluate the Cooper-Frye formula to compute their $p_T$ spectra and $v_2$, as follows.

Utilizing Eq.~(\ref{rrm1}) for a single $c$-quark,
$f_c(\vec x^*,\vec p_2^*)$= $(2\pi)^3\delta^3(\vec p_2^*-\vec p_c^*)/d^3x^*$
and thermal light antiquarks,
$f_{\bar{q}}(\vec x^*,\vec p_1^*)$=$g_{\bar q} e^{-E(\vec p_1^*)/T_H}$,
in a hydro cell at $T_H$, and integrating over the meson
momentum, $\vec p^*$, we obtain
\begin{align}
\label{P_M}
P_M(p_c^*)=P_0\int \frac{d^3\vec p_1^*}{(2\pi)^3} g_q e^{-E(\vec p_1^*)/T_H} \frac{\gamma_M}{\Gamma_M}\sigma(s)v_{\rm rel} \   ,
\end{align}
representing the recombination probability to form a charm meson $M$ from a $c$-quark of momentum $\vec p_c^*$. Likewise, one finds for baryon formation
\begin{align}
\label{P_B}
P_B(p_c^*)&=P_0\int \frac{d^3\vec p_1^*d^3\vec p_2^*}{(2\pi)^6} g_1 e^{-E(\vec p_1^*)/T_H}g_2 e^{-E(\vec p_2^*)/T_H}
\nonumber \\&\times\frac{\gamma_B}{\Gamma_B}\frac{\gamma_{dq}}{\Gamma_{dq}} \sigma(s_{12})v_{\rm rel}^{12}\sigma(s_{d3})v_{\rm rel}^{dq3} \ .
\end{align}
Here, we have introduced an overall normalization, $P_0$, to
require the total recombination probability for a $c$-quark
at rest to be one when summed over all charm-hadron species,
$P_{\rm tot}$($p_c^*$=0)=$\sum_M P_M(0)$+$\sum_B P_B(0)$=1
(with increasing $p_c^*$, $P_{\rm tot}(p_c^*)$ drops off and
``left-over" $c$-quarks will be hadronized via fragmentation~\cite{He:2019tik}).
The pertinent PSDs of hadrons from recombination of a
single $c$ quark are then evaluated by sampling the thermal light-quark PSD as
$f_q(\vec x^*,\vec p_1^*)\sim \sum_n\delta^3(\vec p_1^*-\vec p_{1n}^*)/d^3x^*$, with thermal weights in the fluid
restframe. Using Lorentz invariance of the meson PSD,
$f_M(\vec x,\vec p)=f_M(\vec x^*,\vec p^*)$,
and of $E_M(\vec p^*)\delta^3(\vec p^*-\vec p_{1n}^*-\vec p_c^*)=E_M(\vec p)\delta^3(\vec p-\vec p_{1n}-\vec p_c)$, we
plug $f_{M}(\vec x,\vec p)$ into Eq.~(\ref{dNdp}) and integrate over $\vec p$ to obtain
\begin{align}
\frac{dN_M}{dy}|_{y=0}=
\sum_n\frac{p\cdot d\sigma_H \ \sigma(s)v_{\rm rel}}
{m_M\Gamma_M (d^3x^*)^2}
\equiv \sum_n\Delta N_M[n] \ ,
\end{align}
where we have further exploited boost invariance of
the underlying hydro evolution to convert the space-time rapidity to momentum space rapidity.
For a given $c$-quark and sampling step $n$, the meson momentum, $\vec p=\vec p_{1n}+\vec p_c$, is fully specified, \ie, the
$\Delta N_M[n]$'s form a distribution in $\vec p$ whose sum needs to
recover the recombination probability,
$P_M(p_c^*)$, as given above. The $\Delta N_M[n]$'s are then binned into ($p_T,\phi_p$) histograms to yield the invariant meson spectrum, $dN_M/p_Tdp_Td\phi_pdy$ for a given $\vec p_c$. An analogous procedure is conducted for charm baryons by sampling two static thermal-light quark PSDs.

Our numerical calculations below are carried out
at $T_H$=170\,MeV with resonance widths
$\Gamma_M$$\simeq$0.1\,GeV,
$\Gamma_{dq}$$\simeq$0.2\,GeV and $\Gamma_B$$\simeq$0.3\,GeV,
compatible with the values from the thermodynamic
$T$-matrix~\cite{Riek:2010fk}, as in our previous
work~\cite{He:2011qa}.
We have checked that upon doubling all widths, our final results for the $D$-meson observables change by less than 10\% while
the $\Lambda_c$ suffers additional suppression, by up to
$\sim$30\% at intermediate $p_T$$\simeq$6\,GeV (not included in our uncertainty estimates shown below). However, for large widths the quasiparticle approximation implicit in the current
RRM needs to be replaced by off-shell energy integrals over
spectral functions, which will be deferred to a future study. Also note that we utilize a light diquark as ``doorway state", as the heavy-light color-spin interaction is HQ mass suppressed (in either case, the same equilibrium benchmark for the formed baryon applies,
irrespective of its substructure).
Our charm-hadron spectrum includes all states listed by the PDG
plus additional charm baryons as predicted by the RQM
and lQCD~\cite{He:2019tik}, and essentially figuring in our recent study of 5.02\,TeV $pp$ data for $\Lambda_c$
production~\cite{Acharya:2017kfy}.
Based on this spectrum, the
probability normalization in Eqs.~(\ref{P_M}) and (\ref{P_B})
amounts to $P_0$=3.6. Note that this does not affect the
relative abundances of the various hadrons nor their $p_T$ dependence.
The last ingredient needed to obtain the overall norm of the charm-hadron spectra (equivalent to a $c$ quark fugacity at
$T_H$) is the total charm cross section (which again does not
affect any ratio or $p_T$ dependence).
With $d\sigma_{c\bar c}/dy$$\simeq$1.0\,mb from midrapidity ALICE 5.02\,TeV $pp$ data~\cite{He:2019tik}, a binary nucleon-nucleon collision number of $N_{\rm coll}$$\simeq$1370 and a $\sim$20\% shadowing~\cite{Eskola:2009uj} for
0-20\% $\sqrt{s_{\rm NN}}=5.02$\,TeV Pb-Pb collisions, we obtain $dN_c/dy$$\simeq$15.4.

\begin{figure}[!t]
\hspace{-0.35cm}
\begin{minipage}{0.249\textwidth}
\includegraphics[width=1.18\textwidth]{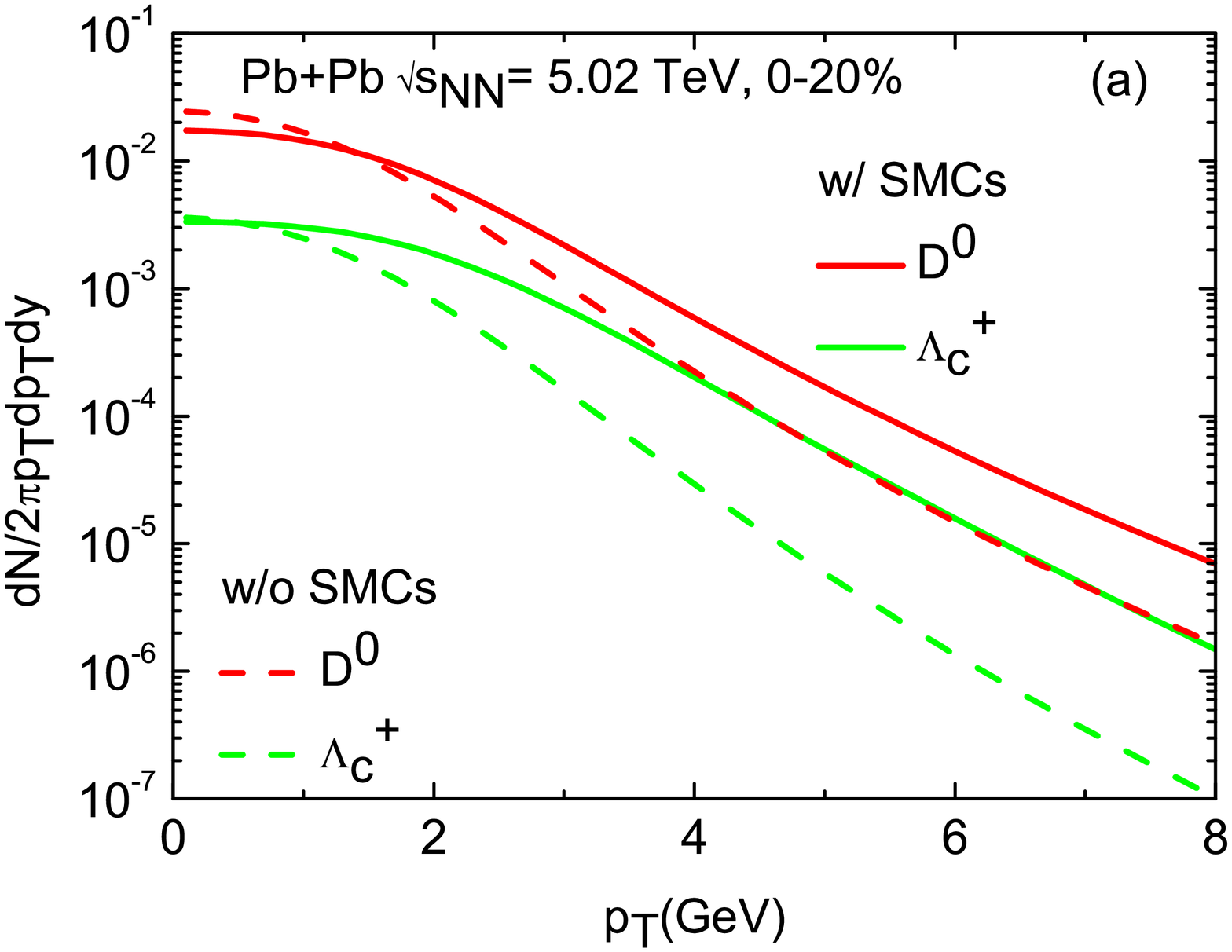}
\end{minipage}
\hspace{-0.22cm}
\begin{minipage}{0.249\textwidth}
\includegraphics[width=1.18\textwidth]{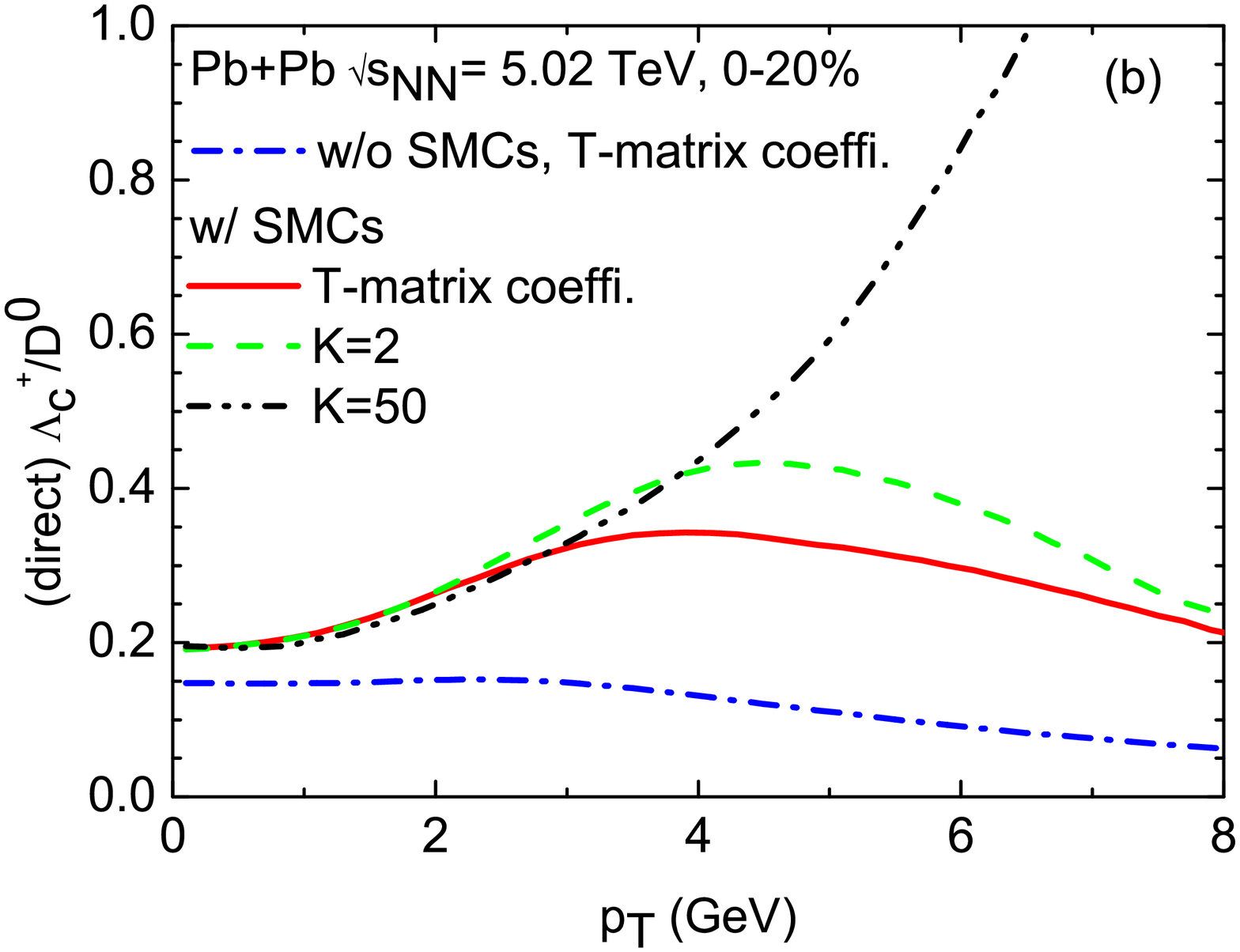}
\end{minipage}
\caption{(Color online) (a): Direct $D^0$ and $\Lambda_c^+$ spectra from hydro-Langevin-RRM simulations with
baseline $T$-matrix $c$-quark thermalization rate, in comparison with the counterparts without SMCs.
(b): The pertinent $\Lambda_c^+/D^0$ ratio with (red line) and without (dash-dotted line) SMCs, and when using a
$K$-factor of 2 (dashed line) and 50 (dash-double dotted line) in the baseline $T$-matrix rate including SMCs.}
\label{fig_directratio_recomprob}
\end{figure}

{\it Direct $\Lambda_c^+/D^0$ ratio from RRM.--}
We now deploy the event-by-event Langevin-RRM simulation with $T$-matrix
transport coefficients in the QGP~\cite{Riek:2010fk}, first
focusing on {\em direct} production of $\Lambda_c$ and $D^0$
(\ie, without feeddown from excited states). The initial
$c$-quark spectra are taken from the FONLL
package~\cite{Cacciari:1998it} as used in our recent study of $\sqrt{s}$=5.02\,TeV $pp$ data~\cite{He:2019tik}.
The resulting RRM-generated spectra of $D^0$ and $\Lambda_c^+$
and their ratio right after hadronization are shown in
Fig.~\ref{fig_directratio_recomprob}, with and without the inclusion of SMCs (the latter scenario corresponds to our
previous implementation~\cite{He:2014cla}, where $c$-quark
conservation was implemented on average, not event by event, \ie, in momentum space only).
The SMCs cause the $D^0$ and $\Lambda_c^+$ spectra to be significantly harder, and the pertinent $\Lambda_c^+/D^0$ ratio
is much enhanced at intermediate $p_T$=3-6\,GeV, relative to their counterparts without SMCs. The key mechanism is relatively fast $c$ quarks moving to the outer parts of the fireball where
they find a higher density of significantly harder light-quark
spectra for recombination, an effect that enters squared for
production of $\Lambda_c$ baryons. Consequently, their RRM yield
toward larger labframe $p_T$ is further enhanced relative to
$D^0$ mesons.

We have numerically verified that in the limit of large
$c$-quark thermalization rates, the {\it absolute} $p_T$
spectra and $v_2$ of $D^0$ and $\Lambda_c^+$ from the event-by-event Langevin-RRM implementation agree well with the direct hydro
calculation (recall Fig.~\ref{fig_eq-map}),
\ie, the ``equilibrium mapping" is maintained in the presence of SMCs. Figure~\ref{fig_directratio_recomprob} also illustrates that, relative to the baseline calculation (solid line), a
$K$-factor of 2 in the HQ thermalization rate enhances the $\Lambda_c/D^0$ ratio only a little, much less than the SMC
effect.

{\it Charm-Hadron Spectra and Ratios.--}
\begin{figure}[!t]
\hspace{-0.35cm}
\begin{minipage}{0.244\textwidth}
\includegraphics[width=1.2\textwidth]{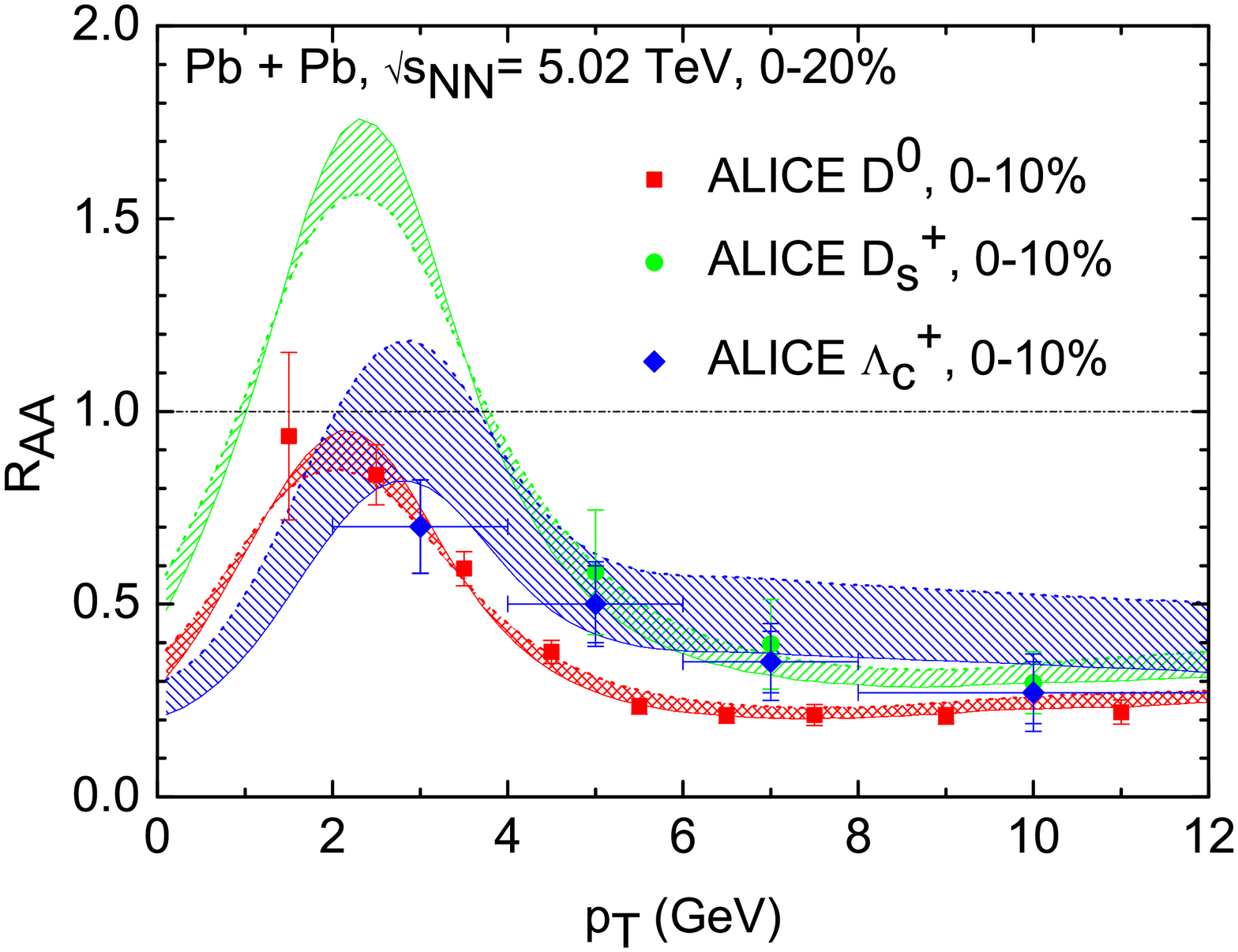}
\end{minipage}
\hspace{-0.22cm}
\begin{minipage}{0.244\textwidth}
\includegraphics[width=1.2\textwidth]{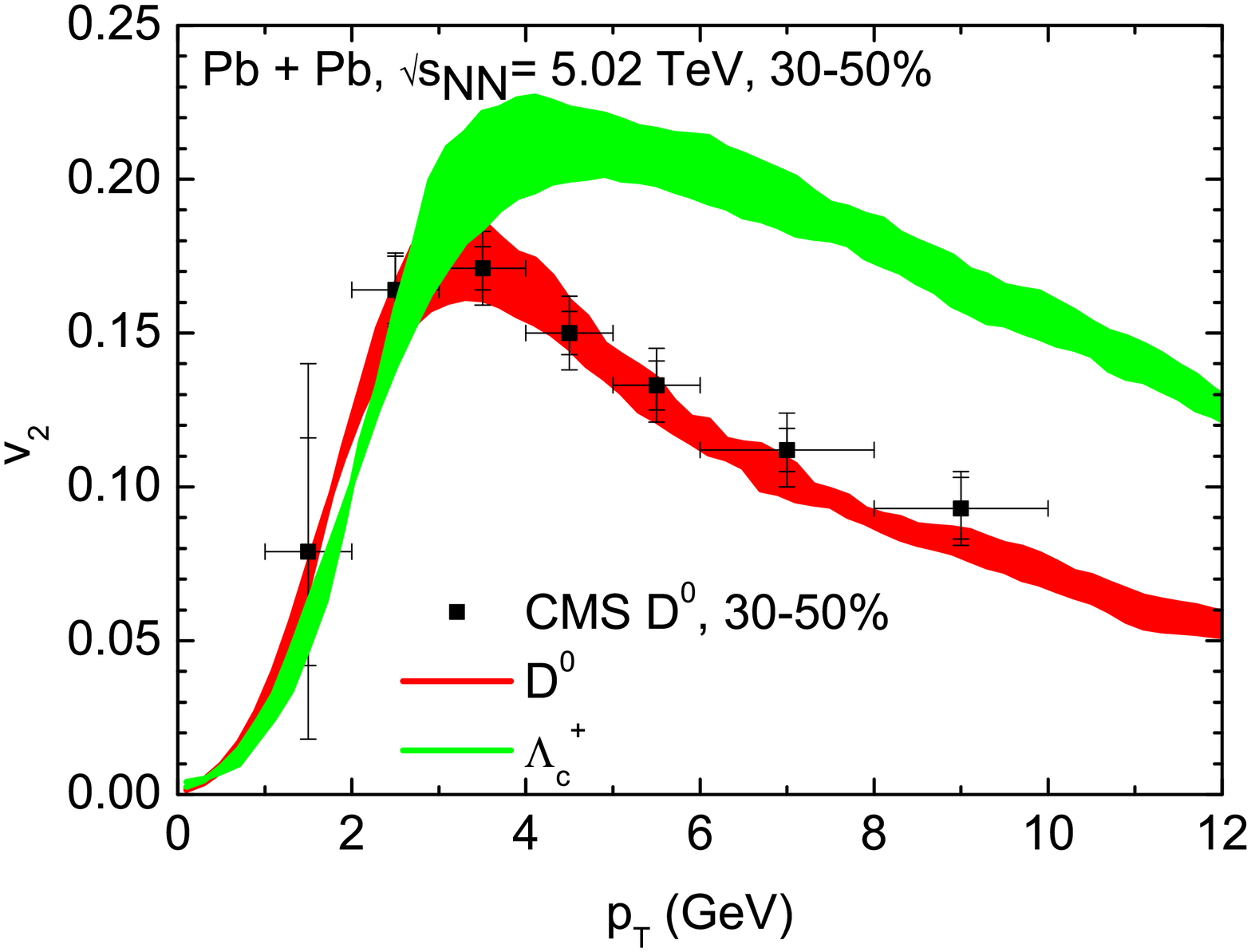}
\end{minipage}
\hspace{-0.22cm}
\begin{minipage}{0.241\textwidth}
\includegraphics[width=1.19\textwidth]{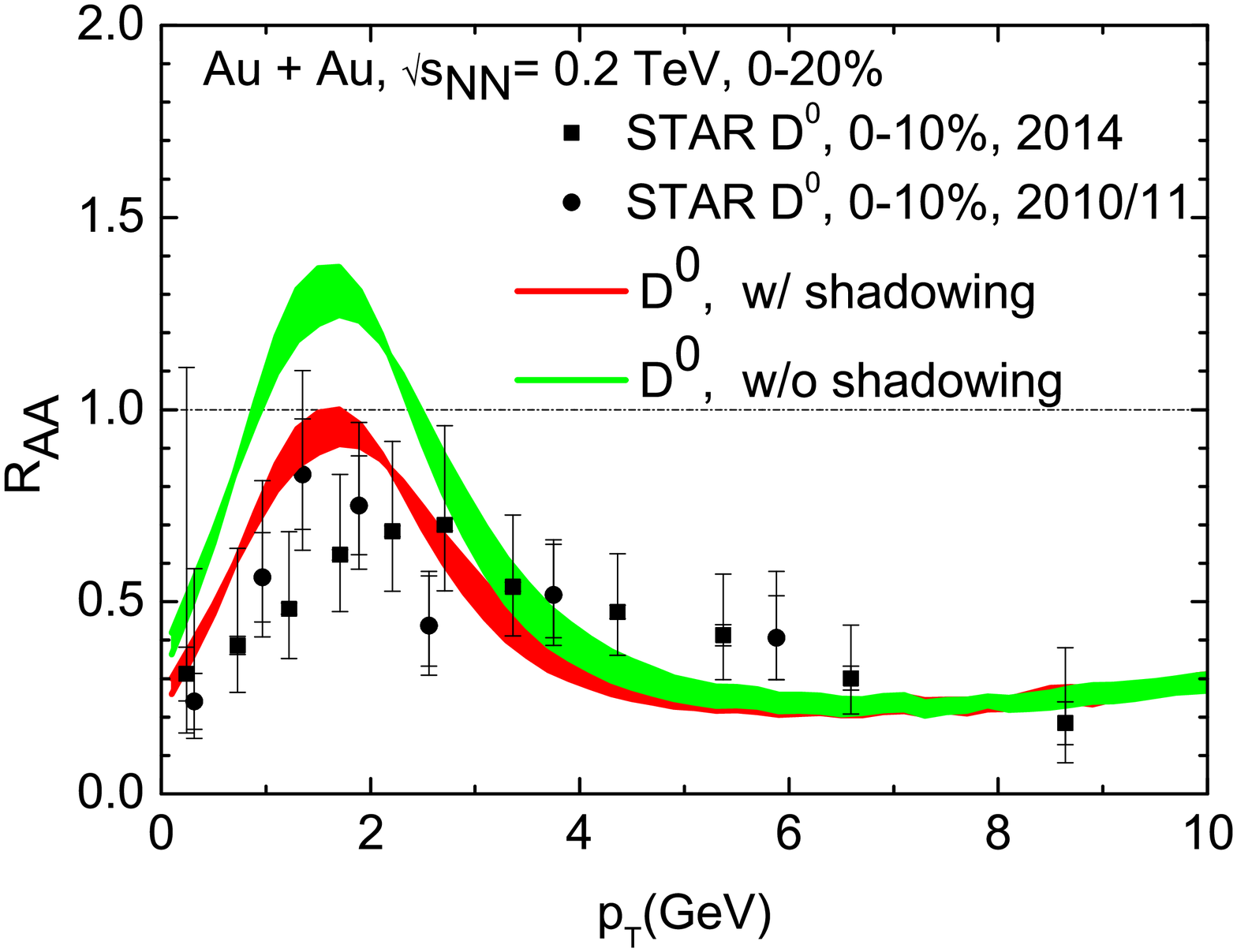}
\end{minipage}
\hspace{-0.22cm}
\begin{minipage}{0.241\textwidth}
\includegraphics[width=1.19\textwidth]{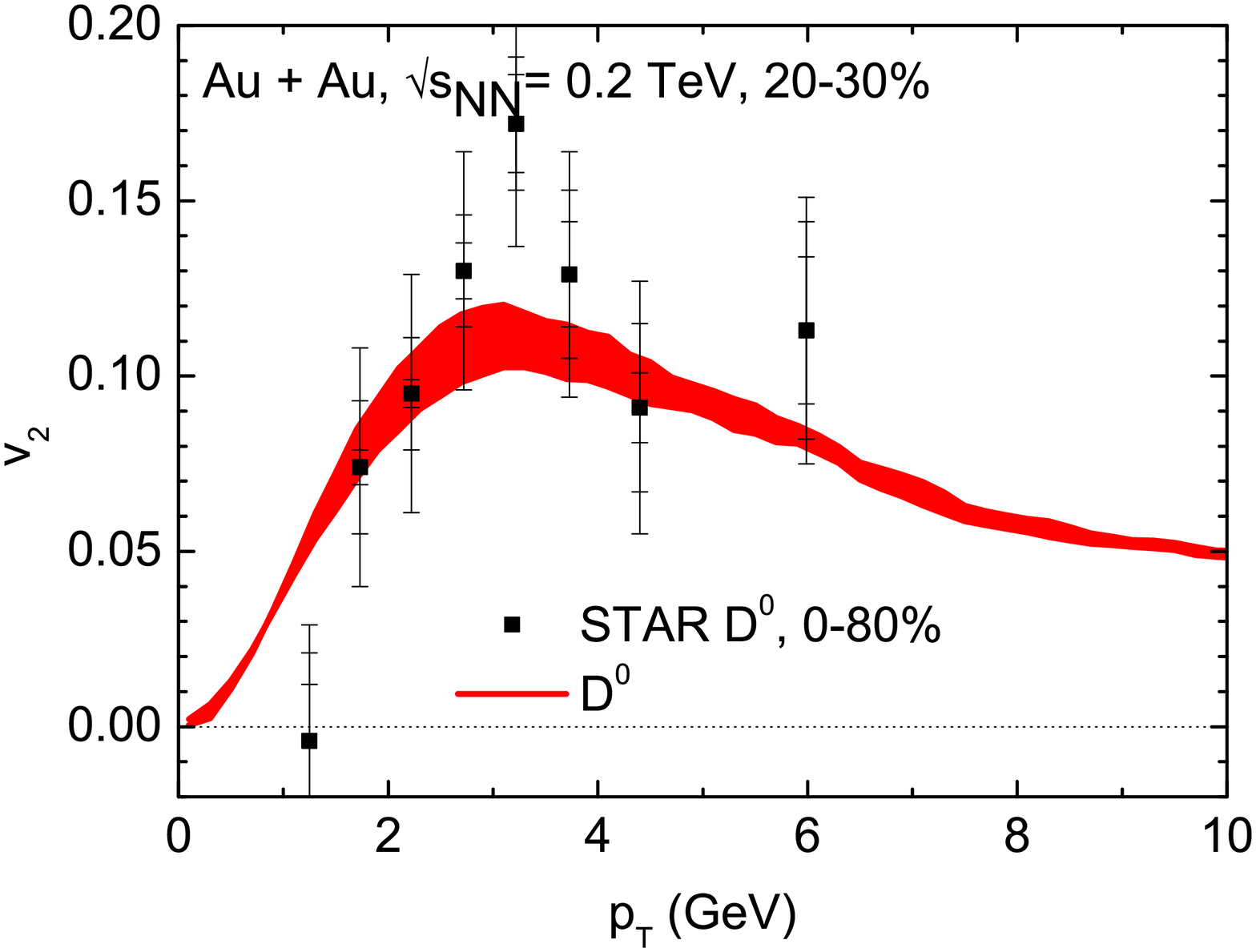}
\end{minipage}
\caption{(Color online) $\RAA$ (left panels) and $v_2$
(right panels) of $D^0$, $D_s^+$ and $\Lambda_c^+$ in
Pb-Pb(5.02\,TeV)(upper panels) and Au-Au(0.2\,TeV) collisions
(lower panels), compared to data~\cite{Acharya:2018hre,Acharya:2018ckj,Adamczyk:2014uip,Adamczyk:2017xur,Sirunyan:2017plt}.
The uncertainty bands for the $\Lambda_c^+$ $\RAA$ are due to
a 50-100\% BR for feeddown from excited states above the $DN$ threshold~\cite{He:2019tik}, and for the other observables due to the effects of hadronic diffusion.}
\label{fig_PbPbD0DsLc-RAA-v2}
\end{figure}
To enable quantitative comparisons of our event-by-event Langevin-RRM
simulations to observables,
we include two further ingredients. First, we continue the Langevin simulations
for all hadrons through the hadronic phase, starting from their PSDs right after hadronization until kinetic freezeout of the
hydrodynamic evolution at $T_{\rm kin}$=110\,MeV (as obtained
from fits to bulk-hadron $p_T$ spectra and $v_2$). For
$D$-mesons we employ our previous thermalization
rates~\cite{He:2011yi}; for charm baryons we (conservatively)
scale the $D$-meson rates with a factor of $E_D(p^*)/E_{\Lambda_c}(p^*)$ to account for their higher
masses; pertinent uncertainties will be illustrated in our plots
below. Second, since our approach currently does not include radiative energy loss, we utilize a temperature- and momentum-independent $K$-factor of 1.6 in the QGP diffusion rate, chosen to improve the overall description of the LHC and RHIC data.

The spectra of all excited states are used to perform decay simulations~\cite{He:2019tik} to obtain the inclusive spectra of
the ground-state $D^0$, $D^+$, $D_s^+$ and $\Lambda_c^+$,
which are then converted into nuclear modification factors
\begin{align}
\RAA(p_T)=\frac{dN_{\rm AA}/dp_T}{N_{\rm coll}dN_{pp}/dp_T} \
 ,
\end{align}
elliptic flows, $v_2(p_T)$, and ratios $D_s^+/D^0$ and $\Lambda_c^+/D^0$. For the $\Lambda_c$ calculations, the main
uncertainty is due to unknown branching ratios (BRs) of excited
states, especially those above the $DN$ threshold which may
not decay into a $\Lambda_c$ final state. As in
Ref.~\cite{He:2019tik}, we illustrate that by a range of
BRs of 50-100\% of these states into $\Lambda_c$ final
states (while keeping the denominator of the $\RAA$ fixed using a fit to the $\Lambda_c$ $pp$ spectrum). Note that their large masses augment the collective
flow effect in their contributions to the $\Lambda_c$ spectra
toward higher $p_T$. For the $D$-meson results, we illustrate uncertainties due to the effects of hadronic diffusion.

\begin{figure}[!t]
\hspace{-0.35cm}
\begin{minipage}{0.249\textwidth}
\includegraphics[width=1.18\textwidth]{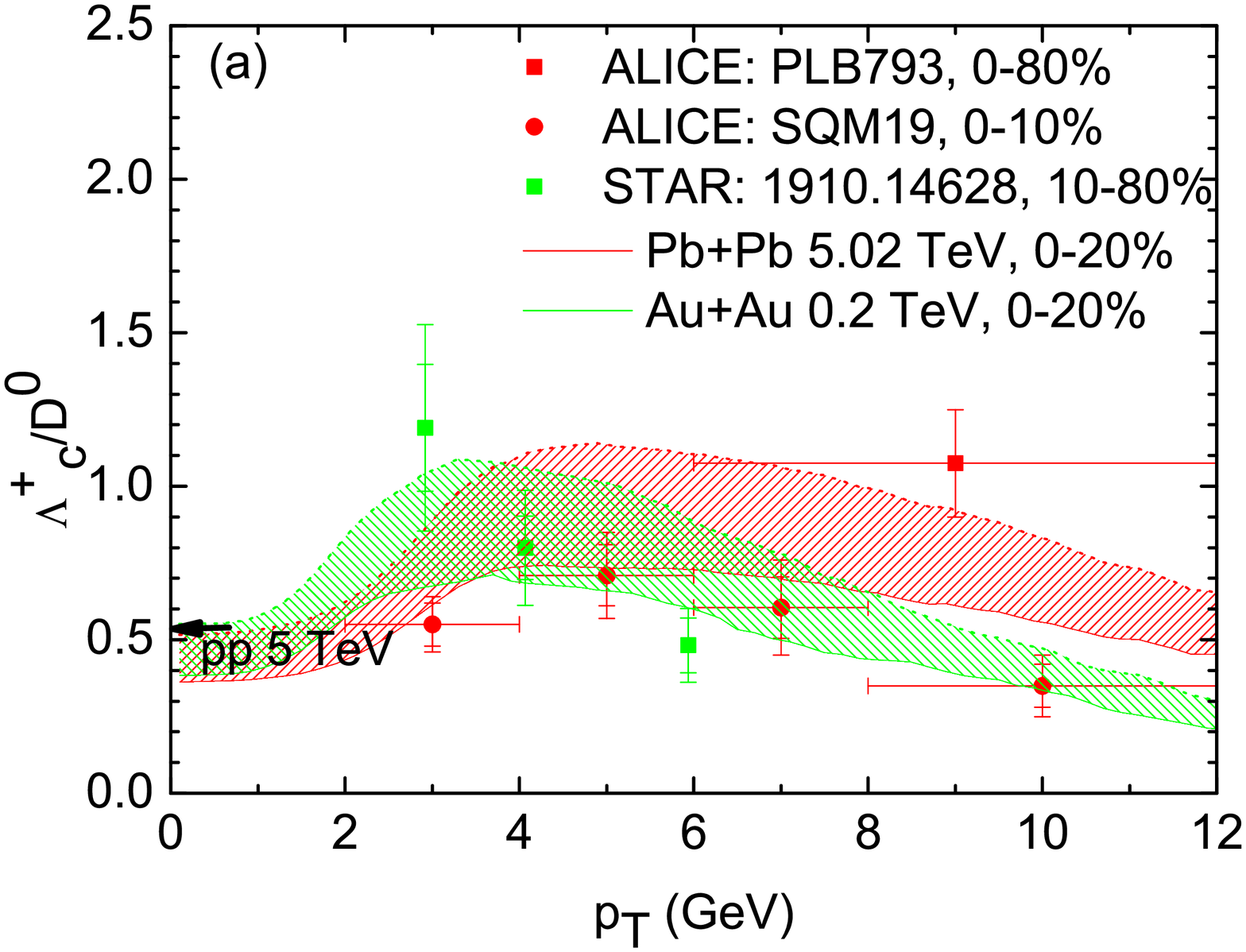}
\end{minipage}
\hspace{-0.22cm}
\begin{minipage}{0.249\textwidth}
\includegraphics[width=1.18\textwidth]{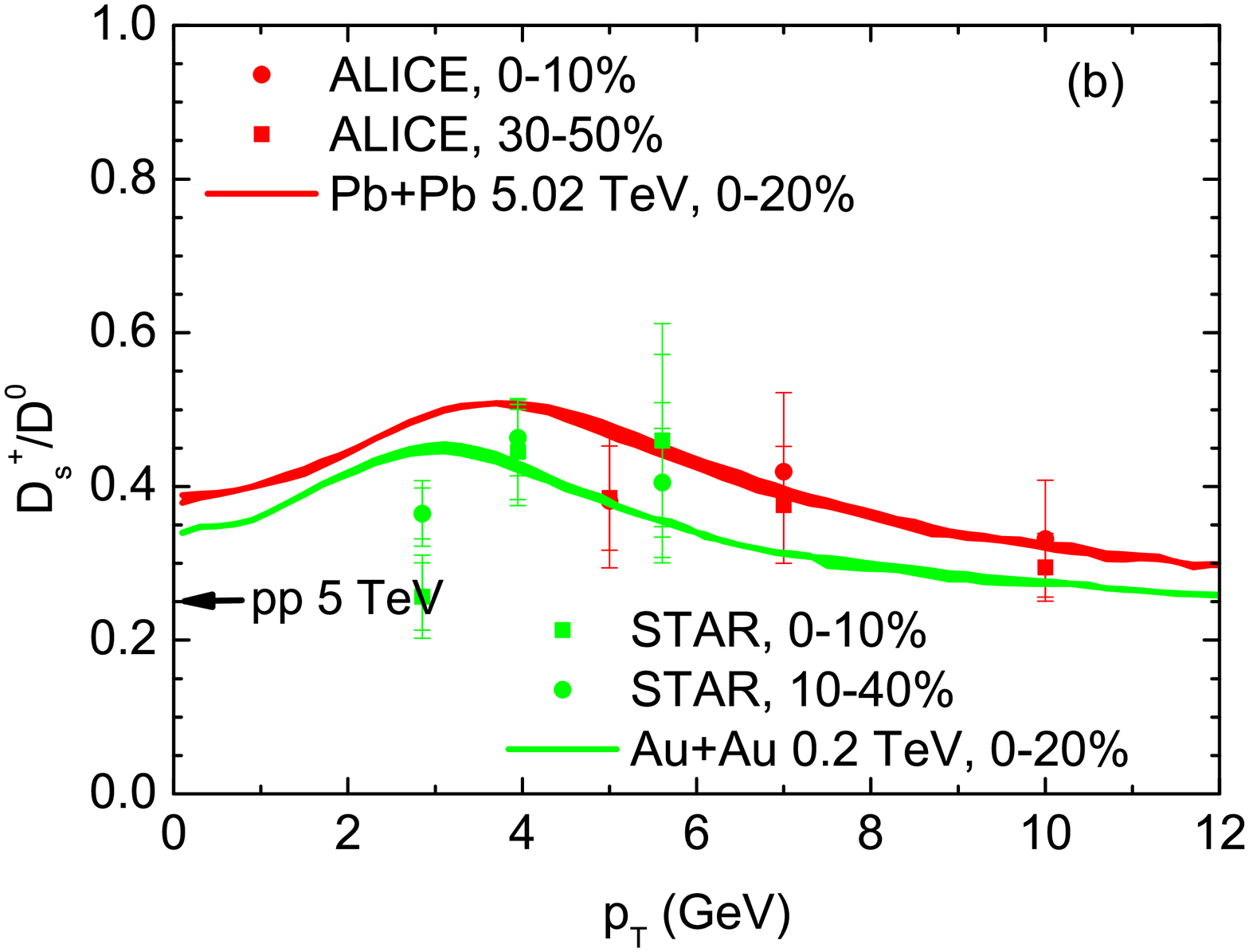}
\end{minipage}
\caption{(Color online) (a): $\Lambda_c^+/D^0$ and (b): $D_s^+/D^0$ ratios compared to
LHC~\cite{Acharya:2018hre,Acharya:2018ckj} and
RHIC~\cite{Zhou:2017ikn,Adam:2019hpq} data.
The uncertainty bands in the $\Lambda_c^+/D^0$ ratios are
due to a 50-100\% BR for $\Lambda_c$ feeddown from excited
states above the $DN$ threshold~\cite{He:2019tik}, and due to hadronic diffusion in the $D_s^+/D^0$ ratio.
The horizontal arrows denote pertinent $pp$ data.}
\label{fig_final-total_LcoverD0-DsoverD0}
\end{figure}

A selection of our results is compared to RHIC and LHC data
in Figs.~\ref{fig_PbPbD0DsLc-RAA-v2} and \ref{fig_final-total_LcoverD0-DsoverD0}.
The suppression hierarchy observed in the $\RAA$ data for
$D^0$, $D_s^+$ and $\Lambda_c^+$ at the LHC is fairly well
reproduced, while the $\Lambda_c/D^0$ ratio tends to be slightly
overpredicted. On the other hand, our results tend to
underpredict this ratio at RHIC toward lower $p_T$. Since the
calculated $\Lambda_c^+/D^0$ ratios approach the chemical
equilibrium limit at low $p_T$, improved data in this regime
at both energies will be very valuable.
Another remarkable consequence of the SMCs in the RRM is a much
improved description of the $D$-meson $v_2$ data out to
higher $p_T$ compared to our previous results.
At RHIC, our results for the $D^0$-$\RAA$, without nuclear shadowing, overestimate the low-$p_T$ STAR
data~\cite{Adamczyk:2014uip} significantly. Assuming a
$\sim$20\% shadowing works better, although most nuclear parton distribution functions do not favor such a scenario. The RHIC results for the
$D^0$ $v_2$ and the $\Lambda_c/D^0$ and $D_s^+/D^0$ ratios are
essentially independent of shadowing.

{\it Summary.--}
In the present paper we have advanced the description of HQ
hadronization in three critical aspects.
First, we developed a 4-momentum conserving recombination
model for baryons,
which is essential for theoretically controlled calculations.
Second, we implemented space-momentum correlations between $c$-quarks and the hydro medium on an event-by-event basis.
Third, we incorporated an improved charm-hadrochemistry,
as previously tested in $pp$ collisions.
We have deployed these developments within our non-perturbative hydro-Langevin-RRM framework, including a moderate $K$ factor in
the QGP diffusion coefficients to simulate hitherto missing contributions from, \eg, radiative interactions.
The new components have significant consequences for the interpretation of RHIC and LHC data, and an ultimately
improved extraction of  HF transport coefficients in QCD matter.
Most notably, the SMCs of fast-moving $c$-quarks with high-flow
partons in the fireball markedly extend the $p_T$ reach of
recombination processes, providing significant enhancements in
$\Lambda_c$ and $D_s$ production at intermediate $p_T$. This
also increases the charm-hadron $v_2$ in this region, in good
agreement with RHIC and LHC $D$-meson data.
Our developments are relevant well beyond the open HF sector in URHICs. We expect the effects of SMCs to shed new light on the
large $v_2$ of $J/\psi$~\cite{Acharya:2017tgv} and light hadrons
at intermediate $p_T$ where current transport and coalescence models tend to underpredict pertinent data. Even for the ``HF
puzzle" in pA collisions, where a large $v_2$ but
$R_{\rm AA}$$\sim$1 is observed~\cite{Dong:2019byy}, the SMCs
could prove valuable, given the explosive nature of the
fireballs conjectured to form in these systems.


{\it Acknowledgments.--}
This work was supported by NSFC grant 11675079 and the U.S.~NSF under grant nos. PHY-1614484 and PHY-1913286.

\end{document}